\documentclass[conference]{IEEEtran}
\IEEEoverridecommandlockouts
\usepackage{cite}
\usepackage{amsmath,amssymb,amsfonts}
\usepackage{algorithmic}
\usepackage{graphicx}
\usepackage{textcomp}
\usepackage{xcolor}
\usepackage{algorithm}
\usepackage{float}
\usepackage{makecell}
\usepackage{subfigure}
\usepackage{booktabs,multirow}
\usepackage{titlesec}
\def\BibTeX{{\rm B\kern-.05em{\sc i\kern-.025em b}\kern-.08em
    T\kern-.1667em\lower.7ex\hbox{E}\kern-.125emX}}
\begin{document}

\setlength{\topmargin}{-0.7in}
\setlength{\columnsep}{0.15 in}

\title{Finite State Machines-Based Path-Following Collaborative Computing Strategy\\for Emergency UAV Swarms\\
\thanks{This work was supported in part by the National Key R\&D Program of China (2023YFC3011502).}
}

\author{\IEEEauthorblockN{Jialin Hu, Zhiyuan Ren, Wenchi Cheng
}
\IEEEauthorblockA{
State Key Laboratory of Integrated Services Networks, Xidian University, Xi'an, China}
E-mail: \{\emph{22011210534@stu.xidian.edu.cn}, \emph{zyren@xidian.edu.cn}, \emph{wccheng@xidian.edu.cn}.
\}
}
\maketitle

\begin{abstract}
Offloading services to UAV swarms for delay-sensitive tasks in Emergency UAV Networks (EUN) can greatly enhance rescue efficiency. Most task-offloading strategies assumed that UAVs were location-fixed and capable of handling all tasks. However, in complex disaster environments, UAV locations often change dynamically, and the heterogeneity of on-board resources presents a significant challenge in optimizing task scheduling in EUN to minimize latency. To address these problems, a Finite state machines-based Path-following Collaborative computation strategy (FPC) for emergency UAV swarms is proposed. First, an Extended Finite State Machine Space-time Graph (EFSMSG) model is constructed to accurately characterize on-board resources and state transitions while shielding the EUN dynamic characteristic. Based on the EFSMSG, a mathematical model is formulated for the FPC strategy to minimize task processing delay while facilitating computation during transmission. Finally, the Constraint Selection Adaptive Binary Particle Swarm Optimization (CSABPSO) algorithm is proposed for the solution. Simulation results demonstrate that the proposed FPC strategy effectively reduces task processing delay, meeting the requirements of delay-sensitive tasks in emergency situations.
\end{abstract}

\section{Introduction}
\indent Unmanned Aerial Vehicles (UAVs) are extensively used in emergency scenarios because of their remarkable scalability and flexibility. In disaster relief, delivering the valuable information collected by UAVs to rescuers is crucial for saving more lives. Additionally, this data often requires further processing before being applied\cite{I_MAS1}. However, individual UAVs have limited resources and cannot independently handle complex tasks\cite{I_MAS2}. Offloading these tasks to cloud servers\cite{I_MAS3} introduces considerable transmission latency, thereby affecting rescue efficiency. Due to advancements in wireless transmission technology, UAVs are now frequently organized into clusters in various fields. Offloading tasks to UAV swarms along their flight path enables efficient transmission and computation to minimize losses through collaborative efforts.\\
\indent Most existing UAV swarm task offloading studies relied on static networks\cite{I_MAS4,I_MAS5,I_MAS6}. For instance, Wang et al.\cite{I_MAS4} conducted a study on a UAV-assisted relay transmission method based on collaborative computational offloading, considering the hovering position of the UAVs as one of the optimization objectives. Guo et al.\cite{I_MAS5} investigated the problem of task scheduling and resource allocation under the constraints of energy consumption of UAVs. Wang et al.\cite{I_MAS6} proposed a scheme for scheduling UAV swarm's computational and communication resources to support Service Function Chains (SFCs). Both assumed that the UAV network topology was fixed. However, disaster sites are complex, fixed-wing UAVs with specific flight trajectories are usually used for wide-area reconnaissance and multi-target search\cite{I_MAS7}. Consequently, the EUN topology usually changes dynamically. Zhao et al.\cite{I_MAS8} investigated a multi-UAV cooperative mobile edge computing system to jointly design UAV trajectories, task assignments, etc. Cheng et al.\cite{I_MAS9} constructed a multi-intelligence cooperative Markov game model to optimize the UAV tasks and power allocation, considering the time-varying position of UAVs.\\
\indent However, all the above works assumed that UAVs were capable of handling all tasks. In a real EUN, UAVs possess heterogeneous on-board resources(e.g., different sensors and image processing units), which results in them only being able to perform specific types of tasks. Wang et al.\cite{I_MAS6} addressed the heterogeneity issue by dividing AI applications into SFCs and scheduling them to multiple UAVs that can create instances. Zheng et al.\cite{I_MAS10} proposed elemental graphs to abstract Virtual Network Functions (VNFs) into independent nodes connected to servers in heterogeneous edge computing networks. The above works were oriented to network service functions and studied the problem of constructing and mapping SFCs, and the networks were all static. However, the tasks in EUNs are diverse, and categorizing task flows and constructing different SFCs to satisfy service requests in dynamic networks will lead to inefficient rescue. Moreover, the nodes can exhibit multiple states triggered by the physical functions, with transitions between these states. Therefore, on-board resources have dependency relationships. Consequently, the current collaborative computing strategies for UAV swarms are not directly applicable to real EUNs, which brings challenges to the scheduling of delay-sensitive tasks and efficient rescue.\\
\indent To solve the above problems, this paper employs a Finite State Machine (FSM) model, which has the characteristics of forced modularity from state definition and facilitates the adaptation to dynamic changes in the environment. Based on this model, a Finite state machines-based Path-following Collaborative computation strategy (FPC) is introduced to optimize the task processing delay. The main contributions of this paper are as follows:
\begin{itemize}
    \item To accommodate the dynamics of EUN while enabling task-oriented resource management, an Extended Finite State Machine Space-time Graph (EFSMSG) model is proposed to describe on-board resources and state transitions, overcoming the heterogeneity of EUN.
\end{itemize}
\begin{itemize}
    \item Considering the requirements of delay-sensitive tasks in emergency scenarios, the FPC strategy is proposed to minimize task processing delays and enable computation while transmitting.
\end{itemize}
\begin{itemize}
    \item Incorporating the cross-time slot mechanism and node function constraints, the Constraint Selection Adaptive Binary Particle Swarm Optimization (CSABPSO) algorithm is proposed to solve the delay optimization problem.
\end{itemize}
\par
\indent The rest of this paper is organized as follows. Section \uppercase\expandafter{\romannumeral2} presents the system model. Section \uppercase\expandafter{\romannumeral3} demonstrates the algorithm in detail. The simulation results and analysis are given in Section \uppercase\expandafter{\romannumeral4}. Finally, Section \uppercase\expandafter{\romannumeral5} concludes the paper. 

\section{System Model and Problem Formulation}
\indent Fig.1 shows a typical architecture for EUN, which generates many delay-sensitive tasks and consists of three layers\cite{I_MAS6}. The first is the ground user, who initiates the task request and receives the result from the upper UAVs within the affected area. The second is a dynamic network composed of UAV formations with fixed flight trajectories. Heterogeneous UAVs from different formations collaborate to provide services for the users. The central UAV in the third layer is responsible for controlling task scheduling and flight trajectory planning, etc.
\vspace{-0.65cm}
\begin{figure}[htbp]
    \centering
    \setlength{\abovecaptionskip}{-0.02cm}
    \setlength{\belowcaptionskip}{-0.5cm}
    \includegraphics[width=0.45\textwidth]{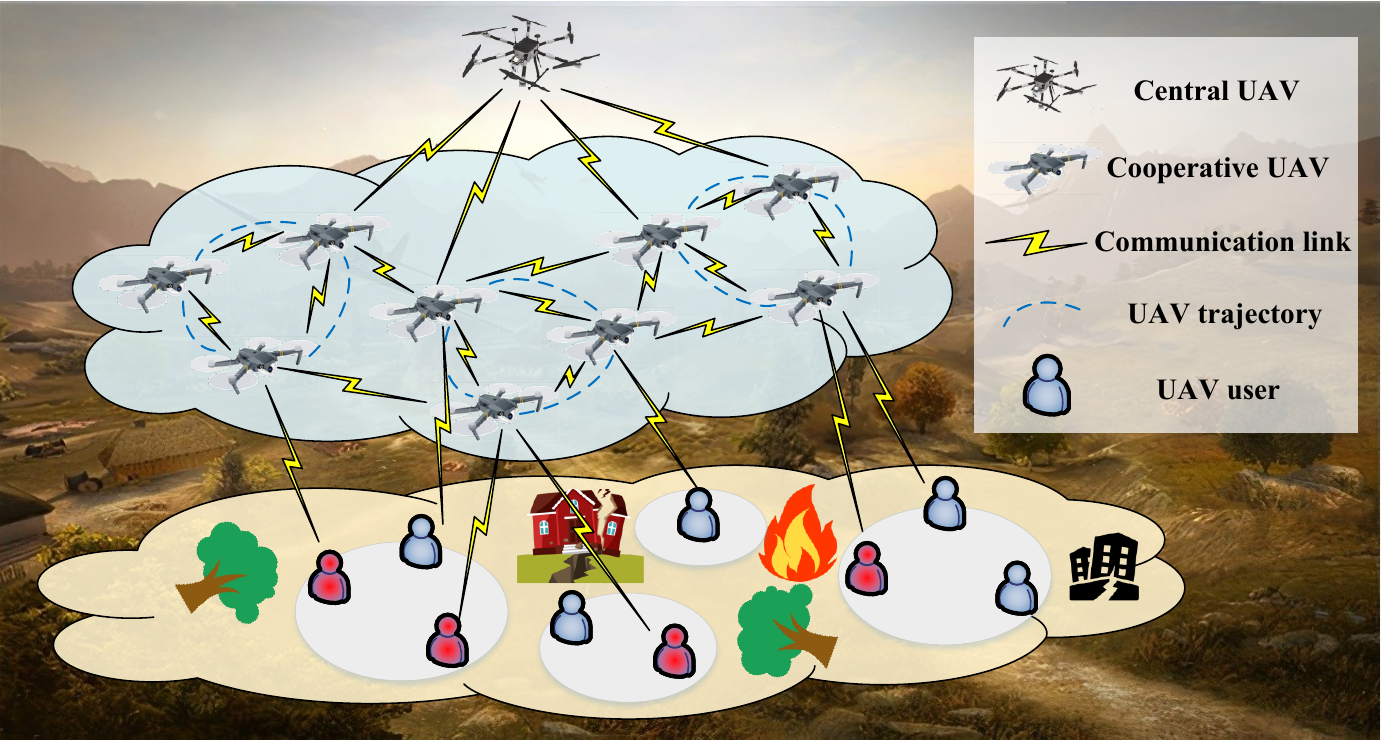}
    \caption{The architecture of EUN.}
    \label{fig1}
\end{figure}
\vspace{-0.4cm}
\subsection{Network Model of EUN}
\indent This subsection uses a Space-time Graph (SG) model to analyze the EUN, which has been widely used in dynamic networks\cite{I_MAS11}. As the SG model relies on known trajectories, the EUN flight trajectory model is constructed. Considering general emergency scenarios, it is assumed that the center UAV hovers at the highest position within the network. The cooperative UAVs are divided into multiple formations, each with a fixed height, and they perform uniform circular motions around distinct centers. To facilitate the modeling, a spatial right-angle coordinate system is established with the center UAV as the coordinate origin, the \textit{xoy} parallel to the ground, and the z-axis perpendicular to the ground. For brevity, all the UAVs mentioned below are cooperative UAVs. $U = \{ {u_1},  \cdots ,{u_p}\} $ is defined as the set of UAVs, and \textit{p} is the number of UAVs. For any ${u_d} \in U$, it can be described by the parameters $\{ {R_d},{\overrightarrow \theta  _d},{\overrightarrow \omega  _d}\} $. Where ${R_d}$ is the flight radius, ${\overrightarrow \omega  _d}$ is the angular velocity, and ${\overrightarrow \theta  _d}$ is the initial phase angle, which is defined as the angle between its line with the center of the circle and the positive direction of the x-axis.\\
\vspace{-0.3cm}
\begin{figure}[h]
    \centering
    \setlength{\abovecaptionskip}{-0.02cm}
    \setlength{\belowcaptionskip}{-0.4cm}
    \includegraphics[width=0.35\textwidth]{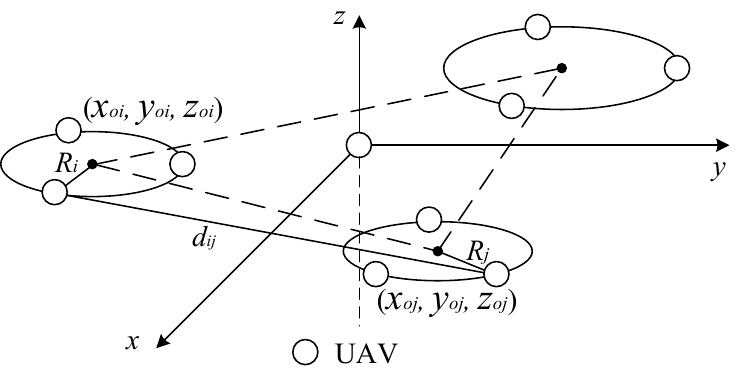}
    \caption{Flight trajectory model of EUN.}
    \label{fig2}
\end{figure}
\vspace{-0.2cm}
\indent As shown in Fig. 2. Where $({x_{oi}},{y_{oi}},{z_{oi}})$ and $({x_{oj}},{y_{oj}},{z_{oj}})$ are the coordinates of the center of the circle of ${u_i}$ and ${u_j}$ respectively, the coordinate of ${u_i}$ can be given by:
\vspace{-0.15cm}
\begin{equation}\label{eq:1}
\vspace{-0.05in}
{x_i}(t) = {x_{oi}} + {R_i}\cos ({\overrightarrow \theta  _i} + {\overrightarrow \omega  _i}t)
\vspace{-0.03in}
\end{equation}
\begin{equation}\label{eq:2}
\vspace{-0.05in}
{y_i}(t) = {y_{oi}} + {R_i}\sin ({\overrightarrow \theta  _i} + {\overrightarrow \omega  _i}t)
\vspace{-0.03in}
\end{equation}
\begin{equation}\label{eq:3}
\vspace{-0.05in}
{z_i}(t) = {z_{oi}}
\vspace{-0.03in}
\end{equation}
\indent Therefore, the euclidean distance ${d_{ij}}$ between any two UAVs at moment \textit{t} can be given by:
\vspace{-0.08cm}
\begin{equation}\label{eq:4}
\vspace{-0.1in}
\resizebox{0.9\hsize}{!}{$\begin{aligned}
{d_{ij}}(t) = \sqrt {{{({x_i}(t) - {x_j}(t))}^2} + {{({y_i}(t) - {y_j}(t))}^2} + {{({z_i}(t) - {z_j}(t))}^2}}
\end{aligned}$}
\vspace{0.08in}
\end{equation}
\indent The UAV-UAV channel is mainly dominated by the line-of-sight (LoS)\cite{I_MAS12}, and it is assumed that the Doppler effects attributable to the UAVs' movement can be compensated in the physical layer with synchronization techniques\cite{I_MAS13}. Then, the transmission loss can be given by:
\vspace{-0.15cm}
\begin{equation}\label{eq:5}
\vspace{-0.05in}
\resizebox{0.8\hsize}{!}{$\begin{aligned}
{L_{ij}}(t) = {(4\pi {d_{ij}}(t)/\lambda )^2}{\xi _{LoS}} = {(4\pi {d_{ij}}(t)f/c)^2}{\xi _{LoS}}
\end{aligned}$}
\vspace{-0.01in}
\end{equation}
where \textit{f} is the carrier frequency, ${\lambda  = c/f}$ is the carrier wavelength, \textit{c} is the velocity of light, ${\xi _{LoS}}$ is the attenuation factors of LoS. On the basis of the free space propagation theorem, the received signal power of the UAV is:
\vspace{-0.05in}
\begin{equation}\label{eq:6}
\vspace{-0.05in}
\resizebox{0.4\hsize}{!}{$\begin{aligned}
P_r^j(t) = P_t^ig_t^ig_r^j/{L_{ij}}(t)
\end{aligned}$}
\end{equation}
where ${P_t^ig_t^i = EIRP}$ is the transmitting power, ${g_r^j}$ is the receiving gain. The signal-to-noise ratio can be given by:
\vspace{-0.05in}
\begin{equation}\label{eq:7}
\vspace{-0.05in}
\resizebox{0.65\hsize}{!}{$\begin{aligned}
SN{R_{ij}}(t) = \frac{{P_r^j(t)}}{{{\sigma ^2}}} = \frac{{P_t^ig_t^ig_r^j}}{{{{(4\pi {d_{ij}}(t)/\lambda )}^2}
{\xi _{LoS}}{\sigma ^2}}}
\end{aligned}$}
\end{equation}
where ${\sigma ^2}$ is the additive white Gaussian noise. When the modulation method is BPSK, the Bit Error Rate (BER) of transmission between UAVs can be given by\cite{I_MAS14}:
\vspace{-0.05in}
\begin{equation}\label{eq:8}
\vspace{-0.05in}
\resizebox{0.45\hsize}{!}{$\begin{aligned}
P_e^{ij}(t) = \frac{1}{2}erfc(\sqrt {SN{R_{ij}}(t)})
\end{aligned}$}
\vspace{-0.03in}
\end{equation}
\indent Therefore, the condition that defines the existence of a link between UAVs is shown below, where ${P_{e0}^j}$ is the demodulation threshold at the receiving end, and ${lin{k_{ij}}(t) = 1}$ indicates that the link is connected and vice versa interrupted.
\vspace{-0.09in}
\begin{equation}\label{eq:9}
\vspace{-0.07in}
lin{k_{ij}}(t) = \left\{ {\begin{array}{*{20}{l}}
{1,}&{P_e^{ij}(t) \le P_{e0}^j}\\
{0,}&{otherwise}
\end{array}} \right.
\vspace{-0.03in}
\end{equation}
\indent According to Shannon formula, the link capacity at moment \textit{t} is shown below, and \textit{B} is the channel bandwidth:
\vspace{-0.05in}
\begin{equation}\label{eq:10}
\vspace{-0.05in}
{R_{ij}}(t) = lin{k_{ij}}(t) \cdot B{\log _2}(1 + SN{R_{ij}}(t))
\end{equation}
\indent The topological period \textit{T} is divided into \textit{n} time slots. The topology is fixed in $\Delta t = T/n$. The transmission delay of unit bit volume is ${\pi _{ij}^k = 1/{R_{ij}}(k)}$. The weighted adjacency matrix of the EUN of the \textit{k}th time slot can be given by:
\vspace{-0.07in}
\begin{equation}\label{eq:11}
\vspace{-0.05in}
{G_k} = \left[ {\begin{array}{*{20}{c}}
0&{\pi _{12}^k}& \cdots &{\pi _{1p}^k}\\
{\pi _{21}^k}&0& \cdots &{\pi _{2p}^k}\\
 \vdots & \vdots & \ddots & \vdots \\
{\pi _{p1}^k}&{\pi _{p2}^k}& \cdots &0
\end{array}} \right]
\vspace{-0.02in}
\end{equation}
\indent Furthermore, the virtual link ${\pi _i^{k(k + 1)} = \Delta t - t_i^k}$ expresses the data cache latency in the same node between two adjacent slots. ${t_i^k}$ is the consumed time of the \textit{i}th node in the \textit{k}th slot. The weighted matrix between two slots can be given by:
\vspace{-0.07in}
\begin{equation}\label{eq:12}
\vspace{-0.05in}
\resizebox{0.8\hsize}{!}{$\begin{aligned}
{G_{k,k + 1}} = \left[ {\begin{array}{*{20}{c}}
{\pi _1^{k(k + 1)}}& \cdots &\infty & \cdots &\infty \\
 \vdots & \ddots & \vdots & \ddots & \vdots \\
\infty & \cdots &{\pi _i^{k(k + 1)}}& \cdots &\infty \\
 \vdots & \ddots & \vdots & \ddots & \vdots \\
\infty & \cdots &\infty & \cdots &{\pi _p^{k(k + 1)}}
\end{array}} \right]
\end{aligned}$}
\vspace{-0.01in}
\end{equation}
\indent Since the UAV flight trajectory is periodic and the adjacency matrix from time slot \textit{n} to time slot 1 also satisfies Eq.(12). The SG model of EUN is given by:
\vspace{-0.05in}
\begin{equation}\label{eq:13}
\resizebox{0.7\hsize}{!}{$\begin{aligned}
G = \left[ {\begin{array}{*{20}{c}}
{{G_1}}&{{G_{1,2}}}&\infty & \cdots &\infty \\
\infty &{{G_2}}&{{G_{2,3}}}& \cdots &\infty \\
 \vdots & \vdots & \vdots & \ddots & \vdots \\
\infty &\infty &\infty & \cdots &{{G_{(n - 1),n}}}\\
{{G_{n,1}}}&\infty &\infty & \cdots &{{G_n}}
\end{array}} \right]
\end{aligned}$}
\end{equation}

\subsection{Model of EFSMSG}
\indent Relying on the SG, the UAV network can be stabilized. However, the heterogeneity of UAVs must be fully considered when formulating a cooperative strategy in EUNs. Therefore, based on SG, FSM is used to characterize the resources and state transitions equipped on UAVs, and the EFSMSG model is constructed in conjunction with the communication topology.\\
\indent FSM is a tool for modeling object behavior. It can effectively describe the sequence of states experienced by the UAV system. The UAV model based on the FSM can be given by:
\vspace{-0.1cm}
\begin{equation}\label{eq:14}
\vspace{-0.05in}
\resizebox{0.45\hsize}{!}{$\begin{aligned}
{S_{UAV}} = \left\{ {Q,E,\delta ,{q_0},O} \right\}
\end{aligned}$}
\end{equation}
where \textit{Q} is the set of all finite states of the UAV. \textit{E} is the set of triggering conditions, ${\delta }$ denotes state transfer, i.e., the process of transferring from one state to another under different triggering conditions, ${{q_0} \in Q}$ is the initial state, and ${O \subseteq Q}$ is the set of termination states. For example, an FSM model of a UAV performing simple image processing tasks\cite{I_MAS1} is shown in Fig. 3. The arrows denote state transfers, and the text on the arrows is state transition triggering conditions.\\
\begin{figure}[h]
    \centering
    \setlength{\abovecaptionskip}{-0.02cm}
    \setlength{\belowcaptionskip}{-0.8cm}
    \includegraphics[width=0.4\textwidth]{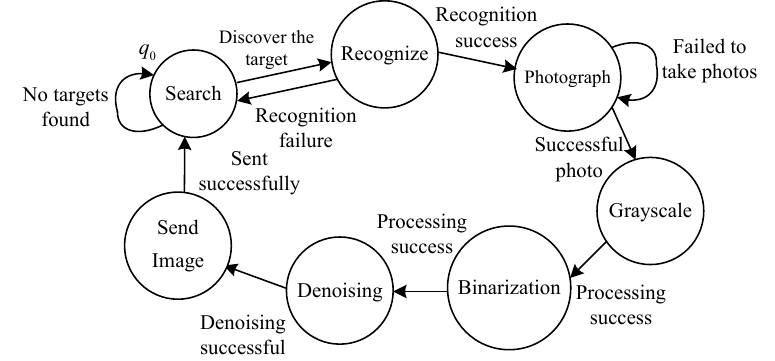}
    \caption{Example of FSM model of UAV.}
    \label{fig3}
    \vspace{-0.7cm}
\end{figure}
\indent Complex tasks usually cannot be accomplished on a single UAV and require multiple UAVs to work together. This process involves not only the state transition of each UAV but also the communication between UAVs. Therefore, this paper considers modeling an EUN as an EFSMSG. e.g., the EFSM model for performing the task of complex image\cite{I_MAS15} is shown in Fig. 4. The communication transfer between UAVs is also considered as a state transition process. Based on the EFSM model and SG model, the EFSMSG model can be constructed as shown in Fig. 5. The model contains the UAV swarm connectivity relationship as well as the state.
\vspace{-0.3cm}
\begin{figure}[h]
    \centering
    \setlength{\abovecaptionskip}{-0.02cm}
    \setlength{\belowcaptionskip}{-0.4cm}
    \includegraphics[width=0.4\textwidth]{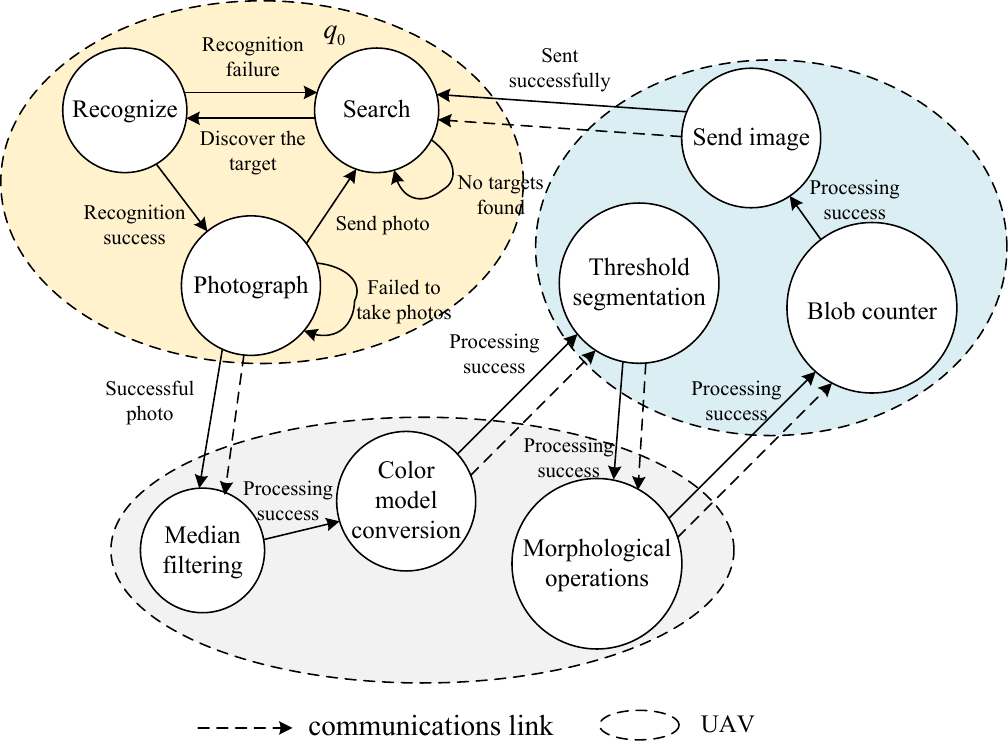}
    \caption{Example of EFSM model of EUN.}
    \label{fig4}
    \vspace{-0.3cm}
\end{figure}
\begin{figure}[t]
    \centering
    \setlength{\abovecaptionskip}{-0.05cm}
    \setlength{\belowcaptionskip}{-0.7cm}
    \includegraphics[width=0.4\textwidth]{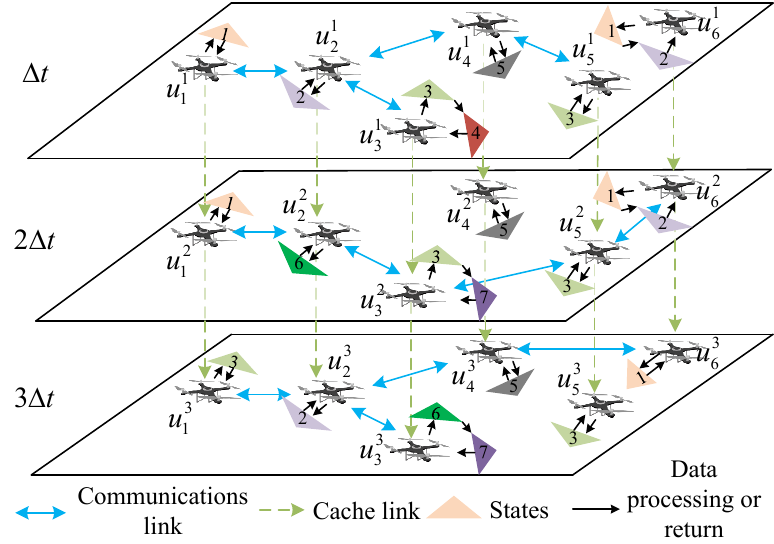}
    \caption{Model of EFSMSG.}
    \label{fig5}
    \vspace{-0.4cm}
\end{figure}

\indent As shown in Fig. 5, for intuition, states are represented as nodes connected to the UAV. The same UAV can transition between multiple states. Assuming there are $N$ types of states in the EUN, and the state of the \textit{l}th type can be denoted as ${f^l} = \left\{ {f_1^l,f_2^l, \ldots ,f_p^l} \right\}$, where $f_j^l(1 \le j \le p)$ represents ${u_j}$ in state $l$. Then all states in the EUN can be given by $F = \left\{ {{f^1},{f^2}, \ldots ,{f^N}} \right\}$. The adjacency matrix of the EUN in the \textit{k}th time slot can be given by:
\vspace{-0.1cm}
\begin{equation}\label{eq:15}
\vspace{-0.05in}
G{p_k} = \left[ {\begin{array}{*{20}{c}}
{{G_k}}&{UF}\\
{FU}&{FF}
\end{array}} \right]
\vspace{-0.01in}
\end{equation}
\vspace{-0.05cm}
\indent ${UF = \left[ {U{f^1},U{f^2}, \cdots ,U{f^N}} \right]}$, ${U{f^l}}$ denotes the delay of the UAV being in state ${{f^l}}$. The \textit{i}th row in the matrix is node ${u_i}$, the \textit{j}th column is $f_j^l$, and the value of the diagonal element is $c{t_{il}}$, which is the latency of the node being in $f_j^l$ to process the unit bit. If the UAV does not have this state, then $c{t_{il}} = \infty $ and the remaining elements are infinity. ${FU = {\left[ {{f^1}U,{f^2}U, \cdots ,{f^N}U} \right]^T}}$, ${{f^l}U}$ denotes the latency for the UAV to invoke ${{f^l}}$ to process the result, and \textit{FF} is the latency for the transition of the UAV state. Since the function of triggering the state is deployed inside the UAV, both are zero and the rest of the elements are infinity. The EFSM matrix of adjacent time slots can be given as follows, with row ${\left[ {{U_k},{f^1}, \cdots ,{f^N}} \right]}$ and column ${\left[ {{U_{k + 1}},{f^1}, \cdots ,{f^N}} \right]}$:
\vspace{-0.2cm}
\begin{equation}\label{eq:16}
\vspace{-0.05in}
\resizebox{0.55\hsize}{!}{$\begin{aligned}
G{p_{k,k + 1}} = \left[ {\begin{array}{*{20}{c}}
{{G_{k,k + 1}}}&\infty & \cdots &\infty \\
\infty &0& \cdots &\infty \\
 \vdots & \vdots & \ddots & \vdots \\
\infty &\infty & \cdots &0
\end{array}} \right]
\end{aligned}$}
\end{equation}
\vspace{-0.2cm}
\indent Then the weighted adjacency matrix of EFSMSG is:
\begin{equation}\label{eq:17}
\vspace{-0.05in}
\resizebox{0.8\hsize}{!}{$\begin{aligned}
Gp = \left[ {\begin{array}{*{20}{c}}
{G{p_1}}&{G{p_{1,2}}}& \cdots &\infty &\infty \\
\infty &{G{p_2}}& \cdots &\infty &\infty \\
 \vdots & \vdots & \ddots & \vdots & \vdots \\
\infty &\infty & \cdots &{G{p_{n - 1}}}&{G{p_{(n - 1),n}}}\\
{G{p_{n,1}}}&\infty & \cdots &\infty &{G{p_n}}
\end{array}} \right]
\end{aligned}$}
\end{equation}
\vspace{-0.3cm}
\subsection{Computation Offloading Based EFSMSG Model}
\indent Based on EFSMSG, this subsection constructs the UAV swarm collaborative computation strategy. Without loss of generality, by representing the task as a Directed Acyclic Graph (DAG) model, the FPC can be transformed into a graph mapping problem from DAG to EFSMSG. Define $\Phi  = (\Psi ,\Gamma )$ as the DAG model of a task. $\Psi  = \left\{ {{\varphi _1}, \ldots ,{\varphi _q}} \right\}$ is the set of sub-tasks, and $\Gamma $ is the logical relationship between sub-tasks. Define ${\Theta _ \uparrow }({\varphi _j}) = \left\{ {\left. {{\varphi _i}} \right|({\varphi _i},{\varphi _j}) \in \Gamma } \right\}$ as the set of forward sub-tasks of ${\varphi _j}$. The task volume ${D_j}$ of ${\varphi _j}$ can be given by:\\
\vspace{-0.25cm}
\begin{equation}\label{eq:18}
\vspace{-0.05in}
\resizebox{0.35\hsize}{!}{$\begin{aligned}
{D_j} = \sum\limits_{{\phi _i} \in {\Theta _ \uparrow }({\phi _j})} {{D_i}{\eta _i}}
\end{aligned}$}
\vspace{-0.05in}
\end{equation}
\vspace{-0.02cm}
\indent ${{\eta _i} \in (0,1]}$ is the data scaling factor, which indicates the change in the amount of data after processing. Definition $B:\Psi  \to F$ is the mapping of nodes, ${\varphi _1}$ maps to the task initiating UAV, ${\varphi _q}$ maps to the result receiving UAV, and the other subtasks map to any node. Considering that the link may be disconnected during transmission, when the data needs to be cached on the UAV until the next time slot, ${{\rho _i}}$ is the number of time slots spanned by the computational subtasks.
\vspace{-0.15cm}
\begin{equation}\label{eq:19}
\vspace{-0.05in}
\hspace{-1mm}
\resizebox{0.9\hsize}{!}{$\begin{aligned}
B\left( {{\varphi _i}} \right) = \left\{ \begin{array}{l}
\left\{ {u_1^1,...,u_1^{1 + {\rho _i}}|1 + {\rho _i} \le n} \right\},{\varphi _i} = {\varphi _1};\\
\left\{ {u_p^k,...,u_p^{k + {\rho _i}}|k \ge 1,k + {\rho _i} \le n} \right\},{\varphi _i} = {\varphi _q};\\
\left\{ {u_d^k,...,u_d^{k + {\rho _i}}|k \ge 1,k + {\rho _i} \le n,1 \le d \le p} \right\},{\rm{otherwise}}{\rm{.}}
\end{array} \right.
\end{aligned}$}
\end{equation}
\vspace{-0.05cm}
\indent Define $Z:\Gamma  \to E$ to denote the mapping of edges ($E$ is the set of edges of EFSMSG). Mapping the edges of DAG to the shortest route between the last node $end(B({\varphi _i}))$ of ${B\left( {{\varphi _i}} \right)}$ to the first node $start(B({\varphi _j}))$ of ${B\left( {{\varphi _j}} \right)}$.
\vspace{-0.1cm}
\begin{equation}\label{eq:20}
\vspace{-0.05in}
Z\left( {\left( {{\varphi _i},{\varphi _j}} \right)} \right) = Pat{h_{end(B({\varphi _i}))start(B({\varphi _j}))}}
\end{equation}
\indent Based on the mapping rule, the latency of the task is shown in Eq. (21). Where ${{T_{comp}}\left( {{\varphi _i}} \right)}$ is the computational delay, ${{T_{accu}}({\varphi _i})}$ is the cumulative delay when proceeding to sub-task ${\varphi _i}$. ${C_{B\left( {{\varphi _i}} \right)}}$ is the computational power of node ${B\left( {{\varphi _i}} \right)}$, and ${{d_{end\left( {B({\varphi _i})} \right)start\left( {B({\varphi _q})} \right)}}}$ is the latency along the shortest path, transmitting unit bit from $\forall {\varphi _j} \in {\Phi _ \uparrow }\left( {{\varphi _i}} \right)$ to ${\varphi _i}$. ${\alpha}$ is the computational complexity of the sub-task, in $cycles/bit$.
\vspace{-0.1cm}
\begin{equation}\label{eq:21}
\vspace{-0.05in}
\hspace{-1mm}
\resizebox{0.9\hsize}{!}{$\begin{aligned}
\begin{array}{l}
T\left( \Phi  \right){\rm{ = }}T\left( {{\varphi _q}} \right) = {T_{comp}}\left( {{\varphi _q}} \right) + {T_{accu}}\left( {{\varphi _q}} \right)\\
{\rm{                     }} = \mathop {\max }\limits_{{\varphi _i} \in {\phi _ \uparrow }({\varphi _q})} [T({\varphi _i}) + {d_{end\left( {B({\varphi _i})} \right)start\left( {B({\varphi _q})} \right)}}{D_i}{\eta _i}] + \frac{{{D_i}{\eta _i}{\alpha _i}}}{{{C_{B({\varphi _q})}}}}.
\end{array}
\end{aligned}$}
\end{equation}
\indent For the same task, there are multiple mapping results that satisfy the rule, which results in different latencies. Therefore, minimizing the task processing latency is to find the optimal mapping strategy. Define the decision matrix \textit{X} as:\\
\vspace{-0.15cm}
\begin{equation}\label{eq:22}
\vspace{-0.05in}
\hspace{-1mm}
\resizebox{0.8\hsize}{!}{$\begin{aligned}
{\bf{X}} = \left[ {\begin{array}{*{20}{c}}
{x\left( {{\varphi _1},u_1^1} \right)}& \cdots &{x\left( {{\varphi _1},u_p^1} \right)}& \cdots &{x\left( {{\varphi _1},u_p^n} \right)}\\
{x\left( {{\varphi _2},u_1^1} \right)}& \cdots &{x\left( {{\varphi _2},u_p^1} \right)}& \cdots &{x\left( {{\varphi _2},u_p^n} \right)}\\
 \vdots & \ddots & \vdots & \ddots & \vdots \\
{x\left( {{\varphi _q},u_1^1} \right)}& \cdots &{x\left( {{\varphi _q},u_p^1} \right)}& \cdots &{x\left( {{\varphi _q},u_p^n} \right)}
\end{array}} \right]
\end{aligned}$}
\end{equation}
\vspace{-0.05cm}
\indent When element $x\left( {{\varphi _i},u_d^k} \right) \in {\bf{X}}$ is equal to 1, it means that sub-task ${\varphi _i}$ is mapped to node $u_d^k$.\\
\vspace{-0.3cm}
\begin{equation}\label{eq:23}
\vspace{-0.05in}
\resizebox{0.6\hsize}{!}{$\begin{aligned}
x\left( {{\varphi _i},u_d^k} \right) \in \left\{ {0,1} \right\},\forall {\varphi _i} \in \Psi ,\forall u_d^k \in {U^T}.
\end{aligned}$}
\end{equation}
\begin{equation}\label{eq:24}
\resizebox{0.5\hsize}{!}{$\begin{aligned}
\sum\limits_{u_d^k \in {U^T}} {x\left( {{\varphi _i},u_d^k} \right)}  = {\rho _i}{\rm{ + 1,}}\forall {\varphi _i} \in \Psi .
\end{aligned}$}
\vspace{-0.05in}
\end{equation}
\indent Therefore, the processing delay of task ${\Phi }$ can be given by:\\
\vspace{-0.15cm}
\begin{equation}\label{eq:25}
\resizebox{1.01\hsize}{!}{$\begin{aligned}
\begin{array}{l}
T\left( {\bf{X}} \right) = \mathop {\max }\limits_{{\varphi _i} \in {\phi _ \uparrow }({\varphi _q})} [T({\varphi _i}) + \sum\limits_{u_d^{k + {\rho _i}},u_h^w \in {U^T}} {{d_{u_d^{k + {\rho _i}}u_h^w}}{D_i}{\eta _i}x\left( {{\varphi _i},u_d^{k + {\rho _i}}} \right)x\left( {{\varphi _q},u_h^w} \right)} ]\\
{\rm{            }} + \sum\limits_{u_h^w \in {U^T}} {\frac{{{D_i}{\eta _i}{\alpha _i}}}{{{C_{u_h^w}}}}x\left( {{\varphi _q},u_h^w} \right)} 
\end{array}
\end{aligned}$}
\vspace{-0.03cm}
\end{equation}
Where ${x\left( {{\varphi _q},u_h^w} \right)}$ denotes the last node of ${{\varphi _q}}$ mapping. Thus, the delay optimization problem can be represented as:
\vspace{-0.1cm}
\begin{equation}\label{eq:26}
\vspace{-0.05in}
\resizebox{0.35\hsize}{!}{$\begin{aligned}
\begin{array}{l}
{\bf{X}} = \arg \min \left( {T({\bf{X}})} \right)\\
s.t.(23),(24)
\end{array}
\end{aligned}$}
\end{equation}
\vspace{-0.3cm}
\section{Algorithm}
\indent Eq. (26) is a multidimensional NP-hard problem with high computational complexity using exact algorithms. Therefore, to solve the latency optimization problem, the CSABPSO algorithm is proposed. Its feasible domain will change adaptively with the increase of the time slot, and the node function constraints are introduced to limit the particle position selection.\\
\indent Assuming the total particle is \textit{M}, and ${I_{\max }}$ is the maximum iteration, i.e. the termination condition of the algorithm. The position and the flying speed of the \textit{m}th particle of the \textit{I}th iteration can be given by:\\
\vspace{-0.25cm}
\begin{equation}\label{eq:27}
\vspace{-0.05in}
\hspace{-4.8mm}
\resizebox{0.95\hsize}{!}{$\begin{aligned}
\begin{array}{l}
X_m^I = size(q,n \times p)\\
{\rm{     }} = \{ \{ x_m^I({\varphi _1},u_1^1), \cdots ,x_m^I({\varphi _1},u_p^n)\} , \cdots ,\{ x_m^I({\varphi _q},u_1^1), \cdots ,x_m^I({\varphi _q},u_p^n)\} \} 
\end{array}
\end{aligned}$}
\end{equation}
\begin{equation}\label{eq:28}
\vspace{-0.05in}
\hspace{-4.8mm}
\resizebox{0.95\hsize}{!}{$\begin{aligned}
\begin{array}{l}
V_m^I = size(q,n \times p)\\
{\rm{    }} = \{ \{ v_m^I({\varphi _1},u_1^1), \cdots ,v_m^I({\varphi _1},u_p^n)\} , \cdots ,\{ v_m^I({\varphi _q},u_1^1), \cdots ,v_m^I({\varphi _q},u_p^n)\} \} 
\end{array}
\end{aligned}$}
\vspace{0.05in}
\end{equation}
Besides, the flying speed of the \textit{m}th particle is updated by Eq. (30). To improve the convergence speed and solution accuracy, the inertia weight ${\mu _I}$ adopts a nonlinear variation strategy:
\vspace{-0.1cm}
\begin{equation}\label{eq:29}
{\mu _I} = {\mu _{start}} - ({\mu _{start}} - {\mu _{end}}){(I/{I_{\max }})^2}
\vspace{-0.05in}
\end{equation}
\vspace{-0.32cm}
\begin{equation}\label{eq:30}
\vspace{-0.05in}
\resizebox{0.8\hsize}{!}{$\begin{aligned}
V_m^{I + 1} = {\mu _I}V_m^I + {\gamma _1}{\beta _1}({p_{{M_{best}}}} - X_m^I) + {\gamma _2}{\beta _2}({g_{best}} - X_m^I)
\end{aligned}$}
\end{equation}
where ${\gamma _1}$ and ${\gamma _2}$ are learning factors, ${{\beta _1}}$ and ${{\beta _2}}$ are random values distributed in the interval [0,1]. The sigmoid function is used to map the particle velocity to the interval [0,1]:\\
\vspace{-0.2cm}
\begin{equation}\label{eq:31}
\vspace{-0.05in}
\resizebox{0.45\hsize}{!}{$\begin{aligned}
S(V_m^{I + 1}) = \frac{1}{{1 + \exp ( - V_m^{I + 1})}}
\vspace{0.08in}
\end{aligned}$}
\end{equation}
\indent The position of each particle ${X_m^{I + 1}}$ represents a mapping strategy, and from Section \uppercase\expandafter{\romannumeral2}, it is clear that subtasks can only be mapped to UAV nodes where the processing state of the subtask exists. Assuming that ${{u_d^k}}$ exists ${{{\varphi _i}}}$ the processing state ${{f_d^i}}$ is denoted as ${ismember({\varphi _i},u_d^k) = 1}$. Therefore, the particle position update formula is as follows:
\vspace{-0.1cm}
\begin{equation}\label{eq:32}
\hspace{-1mm}
\resizebox{0.83\hsize}{!}{$\begin{aligned}
x_m^{I + 1}({\varphi _i},u_d^k) = \left\{ {\begin{array}{*{20}{l}}
{1,}&\begin{array}{l}
S(V_m^{I + 1})({\varphi _i},u_d^k) \ge rand(){\rm{ }}\\
\& ismember({\varphi _i},u_d^k) = 1
\end{array}\\
{0,}&{otherwise}
\end{array}} \right.
\end{aligned}$}
\vspace{-0.05in}
\end{equation}
${S(V_m^{I + 1})}$ is the probability that ${x_m^{I + 1}}$ is 1 and ${rand()}$ is a random number in the interval [0,1]. Then the fitness value can be calculated by the following equation:
\vspace{-0.1cm}
\begin{equation}\label{eq:33}
f(X_m^{I + 1}) = T(X_m^{I + 1})
\vspace{-0.4cm}
\end{equation}
\vspace{-0.2cm}
\begin{table}
    \label{tab:da16}
    \centering
    \begin{tabular*}{\hsize}{@{}@{\extracolsep{\fill}}l@{}}
    \toprule 
    \textbf{Algorithm 1} CSABPSO algorithm \\
    \midrule 
    \textbf{Input:} input a particle swarm with \textit{M},${I_{\max }}$,${{\beta _1}}$,${{\beta _2}}$,${\gamma _1}$,${\gamma _2}$,${{\mu _{start}}}$,${{\mu _{end}}}$;\\
    \ 1: \textbf{for} each particle $m \le M$ \textbf{do}\\\ 2:\quad\quad Initialize ${X_m^{0}}$,${V_m^{0}}$;\\\ 3:\quad\quad Evaluate the fitness function of particle \textit{m} using (33);\\\ 4:\quad\quad Set the current position as the best position of particle ${{p_{{M_{best}}}}}$;\\\ 5: \textbf{end for}\\\ 6: Choose the particle position with the best fitness of all particles as the\\ \quad\ \! global best position ${{g_{best}}}$;\\\ 7: \textbf{for} generation 1 to ${I_{\max }}$ \textbf{do}\\\ 8:\quad\quad \textbf{for} each particle $m \le M$ \textbf{do}\\\ 9:\quad\quad\quad\quad Update the inertia weight ${\mu _I}$ using (29);\\10:\quad\quad\quad\quad Update the velocity and position of particle \textit{m} using (30-32);\\11:\quad\quad\quad\quad Evaluate the fitness function with the new position using (33);\\12:\quad\quad\quad\quad \textbf{if} ${f(X_m^{I + 1}) < f({p_{{M_{best}}}})}$ \textbf{then}\\13:\quad\quad\quad\quad\quad\quad ${{p_{{M_{best}}}} = X_m^{I + 1}}$;\\14:\quad\quad\quad\quad \textbf{end if}\\15:\quad\quad\quad\quad \textbf{if} ${f({p_{{M_{best}}}}) < f({g_{best}})}$ \textbf{then}\\16:\quad\quad\quad\quad\quad\quad ${{g_{best}} = {p_{{M_{best}}}}}$;\\17:\quad\quad\quad\quad \textbf{end if}\\18:\quad\quad \textbf{end for}\\19: \textbf{end for}\\\textbf{Output:} ${f({g_{best}})}$;\\
    \bottomrule
    \end{tabular*}
    \vspace{-0.5cm}
\end{table}

\section{Simulation Results}
\indent To verify the effectiveness of FPC, extensive experiments are presented. The simulation platform is MATLAB. Four UAV formations are set up. Each formation has four UAVs. Referring to \cite{I_MAS1,I_MAS7,I_MAS8,I_MAS16,I_MAS17}, and the rest of the simulation parameters are shown in Table 1. The parameters of CSABPSO are set as: the number of particles ${M = 500}$, ${I_{\max }}=1000$, ${\gamma _1} = {\gamma _2} = 1$, ${\mu _{start}}=1.5$, and ${\mu _{end}}=0.5$.\\
\vspace{-0.4cm}
\begin{table}[h]
  \centering
  \caption{System Parameters of Eun}
  \resizebox{8.8cm}{!}{
    \begin{tabular}{|c|c|c|c|}
    \hline
    Parameter & Value & \makecell[c]{Parameter} & \makecell[c]{Value} \\
    \hline
    $({x_{oi}},{y_{oi}})$ & randomly in 4*4${\rm{k}}{{\rm{m}}^2}$ area & $({z_{oi}})$ & [20, 100]m \\
    \hline
    $\overrightarrow {{\omega _i}} $ & [0.05, 0.8]rad/s & ${R_i}$ & [40, 160]m \\
    \hline
    ${\xi _{LoS}}$ & 3 dB & \textit{f} & 2.4 GHz \\
    \hline
    EIRP & 20 dBm & ${g_r}$ & 3 dB \\
    \hline
    $B$ & 20 MHz & ${\sigma ^2}$ & -100 dBm \\
    \hline
    ${P_{e0}}$ & ${10^{ - 5}}$ & ${C_i}$ & [500, 1200]MHz \\
    \hline
    $\Delta t$ & 1s & ${\eta _i}$ & 0.8 \\
    \hline
    \end{tabular}}
  \label{tab:summary}
\end{table}
\vspace{-0.2cm}
\indent Fig. 7 explores the relationship between task computational complexity and processing delay. When EUN performs the task in Fig. 6\cite{I_MAS1}, the input data size is set to range from 1 Mb to 5 Mb because the images have different resolutions and formats. The task computational complexity factor ranges from [100, 200]\textit{cycle/bit}\cite{I_MAS8}.\\
\begin{figure}[h]
    \centering
    \setlength{\abovecaptionskip}{-0.02cm}
    \setlength{\belowcaptionskip}{-0.8cm}
    \includegraphics[width=0.4\textwidth]{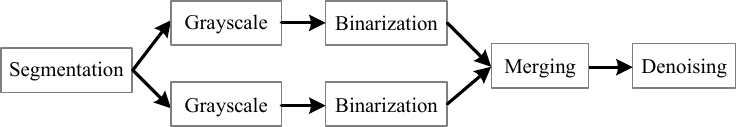}
    \caption{Image processing task model.}
    \label{fig6}
    \vspace{-0.55cm}
\end{figure}
\indent From Fig. 7, it can be seen that with the same amount of data, the delay increases with the increase of computational complexity. Moreover, the impact of different input data sizes is different. For example, when the input data size is 1Mb, the total latency rises smoothly from 0.2375s to 0.4848s with the increase in computational complexity. However, in the case of 5Mb, the total latency jumps at computational complexity 150 to 160. This is because when the data size is small, the task does not need to be processed across time slots, so the increase in computational complexity only makes the computational latency increase, and thus the total latency rises smoothly. However, when the data size is large, the increase in computational delay causes the task to be difficult to complete within the current time slot, at which point the data has to be cached to the next time slot for computation, which causes the transmission delay of the task to rise significantly, and thus the total latency shows a jump in growth.\\
\vspace{-0.5cm}
\begin{figure}[h]
    \centering
    \setlength{\abovecaptionskip}{-0.02cm}
    \setlength{\belowcaptionskip}{-0.2cm}
    \includegraphics[width=0.4\textwidth]{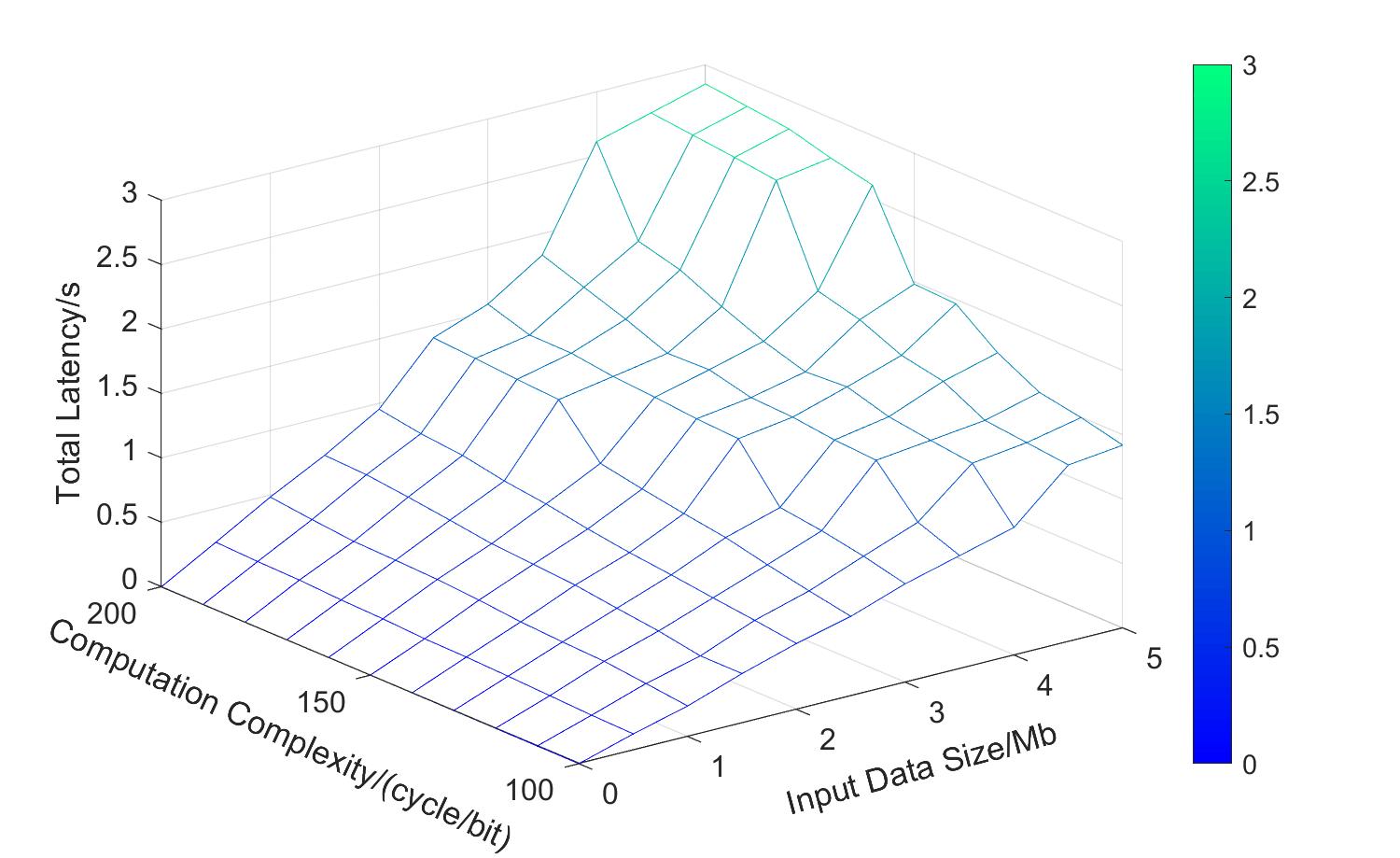}
    \caption{Latency performance for different computational complexities.}
    \label{fig7}
\end{figure}
\vspace{-0.2cm}
\indent For the remote sensing image processing\cite{I_MAS15} and fire detection and tracking\cite{I_MAS18} tasks in real emergency scenarios, Fig. 9 compares the latency performance of cloud computing, local computing, and FPC. The DAG model and computational complexity of the two tasks are shown in Fig. 8.\\
\vspace{-0.6cm}
\begin{figure}[h]
  \centering
  \setlength{\abovecaptionskip}{-0.02cm}
  \subfigure[Remote sensing image(${\alpha}$=100)]{\includegraphics[width=0.3\textwidth]{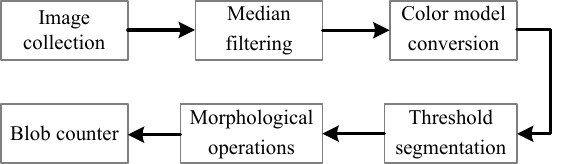}\label{fig:subfig1}}
  \hspace{0.5cm}
  \subfigure[Fire detection tracking(${\alpha}$=200)]{\includegraphics[width=0.4\textwidth]{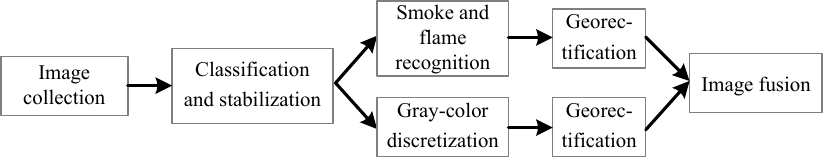}\label{fig:subfig2}}
  \caption{Actual emergency task DAG model.}
  \label{fig8}
\end{figure}

\vspace{-0.3cm}
\indent As can be seen from Fig. 9, the cloud computing curve is higher than the local computing and FPC as the input data size increases. The reason is that the cloud computing center is too far away, which leads to a linear increase in its transmission latency. In addition, the local computing latency is relatively lower than cloud computing but higher than FPC, which is due to the limited computing capability of user terminals, which will cause higher computing latency. FPC offloads tasks to heterogeneous UAV swarms in the transmission path, which not only avoids the long-distance transmission of data but also unites the computing resources of UAV swarms, which reduces the computational latency. When the input data size is 5Mb, the latency performance of FPC improves by 71.63${\% }$ and 60.00${\% }$ compared with cloud computing and local computing. Therefore, FPC is suitable for delay-sensitive tasks in EUN.\\
\vspace{-0.5cm}
\begin{figure}[h]
  \centering
  \setlength{\abovecaptionskip}{-0.01cm}
  \setlength{\belowcaptionskip}{-0.1cm}
  \subfigure[Remote sensing image(${\alpha}$=100)]
  {\label{fig:subfig1}\includegraphics[width=0.49\linewidth]{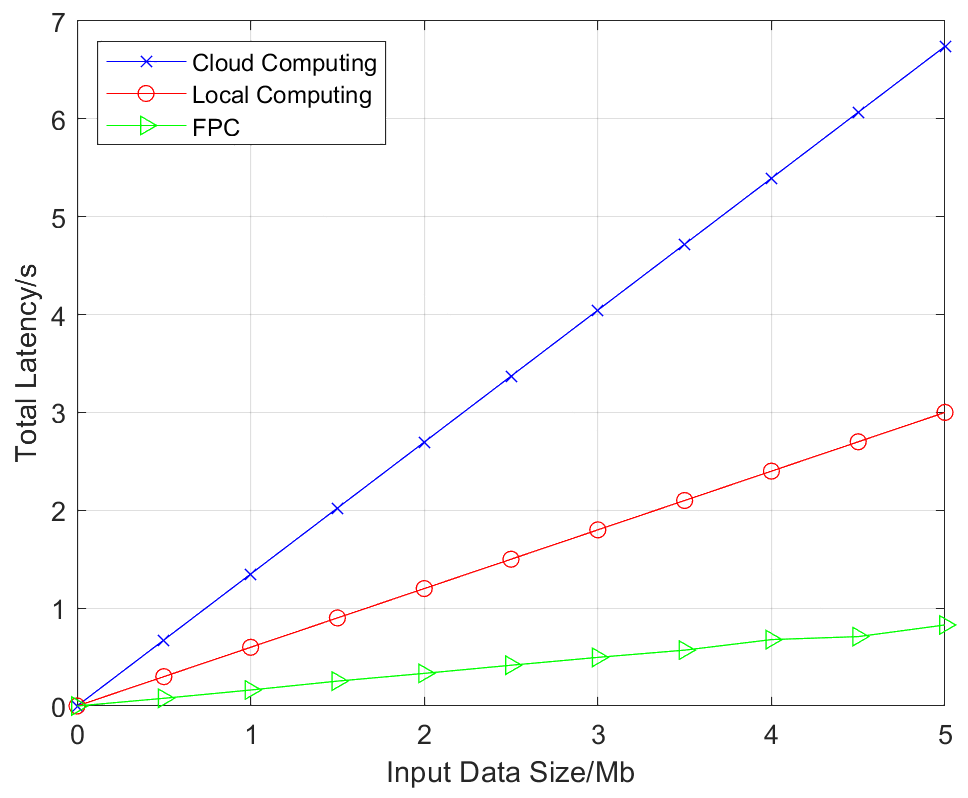}}
  \subfigure[Fire detection tracking(${\alpha}$=200)]
  {\label{fig:subfig2}\includegraphics[width=0.49\linewidth]{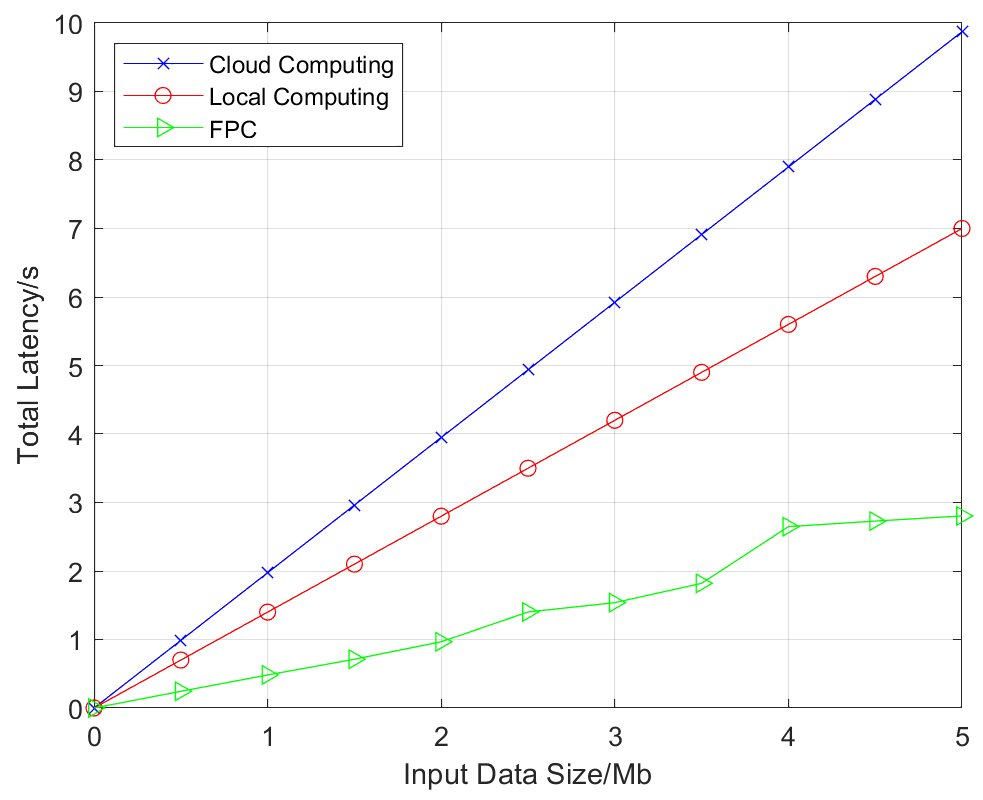}}
  \caption{Latency performance of three computation architectures.}
  \label{fig:subfig_1} 
\end{figure}
\vspace{-0.2cm}
\indent Fig. 10 compares the latency performance of the CSABPSO algorithm with typical load balancing algorithms, i.e., Weighted Rotation (WRR), Greedy Load Balancing (Greedy-LB), Pick-KX. It can be found that the difference between the four algorithms is not significant when the amount of data is small, but when the input data size increases, the latency of CSABPSO decreases significantly. For example, when EUN performs the remote sensing image processing task, under the condition that the input data size is 5Mb, the latency of CSABPSO, WRR, Greedy-LB, and Pick-KX are 0.8282s, 1.3619s, 1.1241s, and 1.4174s. The latency of CSABPSO compared to the remaining three algorithms decreases by 39.19${\% }$, 26.32${\% }$, and 41.57${\% }$ respectively. This is because the mapping strategy of WRR is based on the computational capability of the UAV, while Greedy-LB and Pick-KX are based on the load condition of the UAV, and all three ignore the effect of the transmission link. CSABPSO has a relatively high delay performance by considering the computational capability and the communication capability together.\\
\vspace{-0.8cm}
\begin{figure}[h]
  \setlength{\abovecaptionskip}{-0.01cm}
  \setlength{\belowcaptionskip}{-0.1cm}
  \centering
  \subfigure[Remote sensing image(${\alpha}$=100)]
  {\label{fig:subfig1}\includegraphics[width=0.49\linewidth]{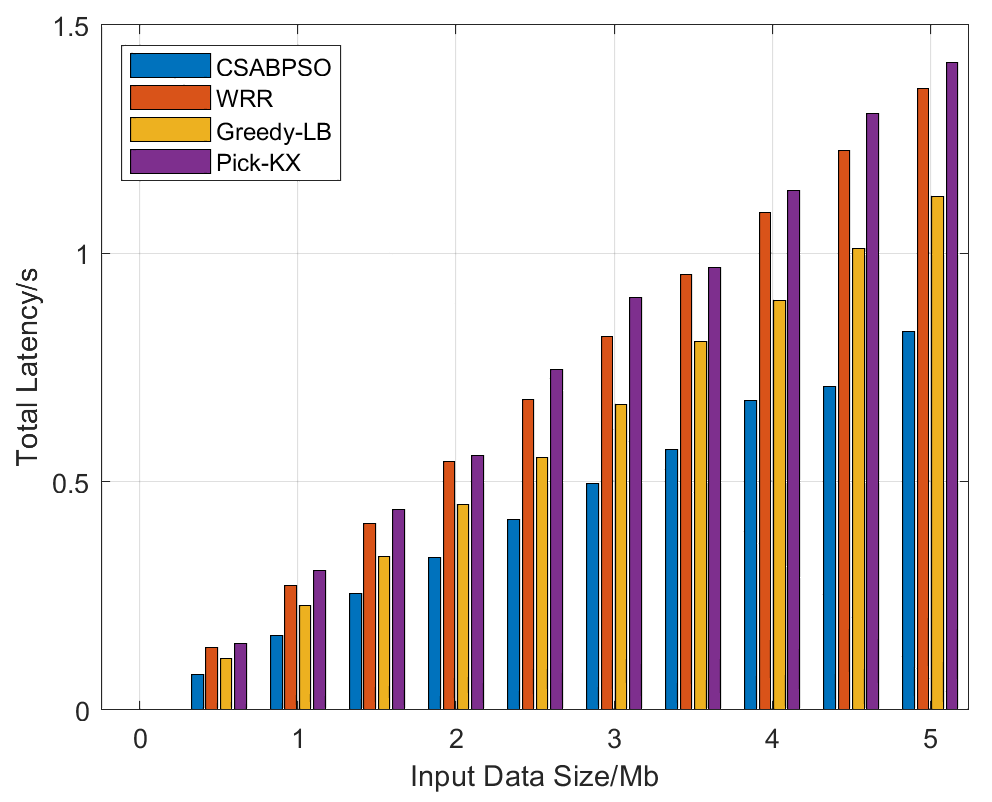}}
  \subfigure[Fire detection tracking(${\alpha}$=200)]
  {\label{fig:subfig2}\includegraphics[width=0.49\linewidth]{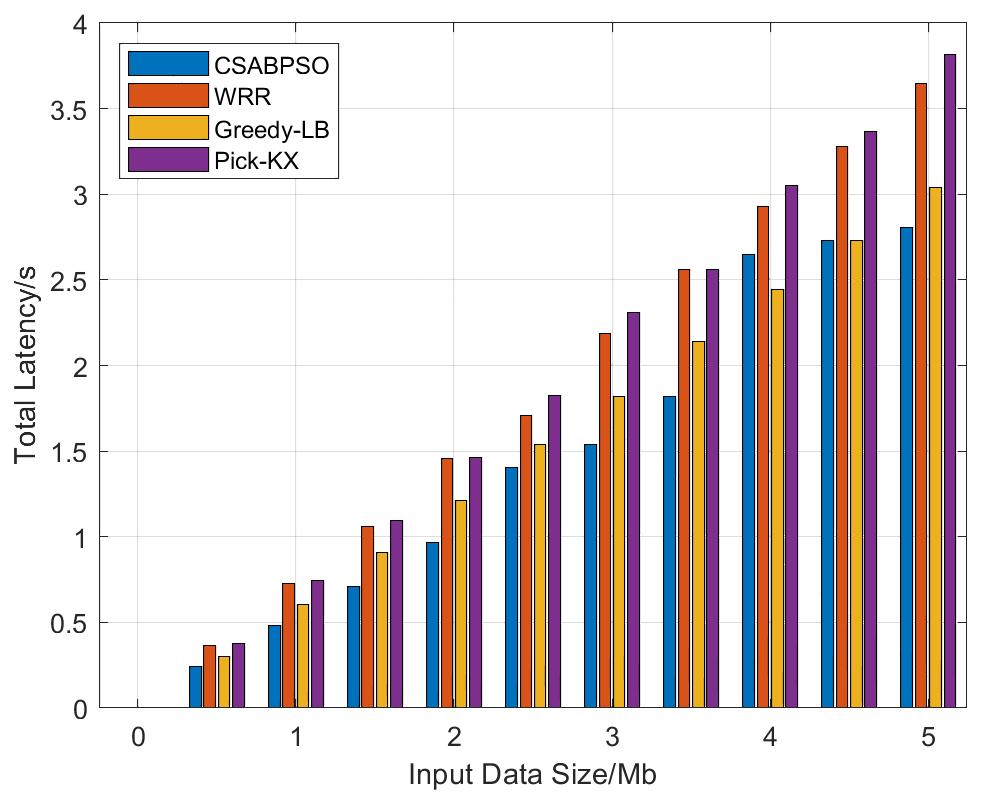}}
  \caption{Latency performance of different algorithms.}
  \label{fig:subfig_1} 
\end{figure}

\vspace{-0.4cm}
\section{Conclusions}
\indent In this paper, we proposed a finite state machines-based path-following collaborative computation strategy for emergency UAV swarms. First, the EFSMSG model was constructed to achieve accurate characterization of on-board resources and state transitions while shielding the dynamics of the EUN. Second, the mathematical model of the FPC strategy was constructed to transform the collaborative computation into the mapping from DAG to EFSMSG. Finally, the CSABPSO algorithm was proposed for solving the problem. Simulation results showed the effectiveness of our proposed FPC and algorithm in terms of latency.

\end{document}